\documentclass[a4paper,oneside,preprintnumbers]{revtex4}
\usepackage{axodraw}
\usepackage[english]{babel}
\usepackage{amsmath}
\usepackage[charter,cal=cmcal,greekuppercase=italicized]{mathdesign}
\usepackage{dsfont}
\usepackage[T1]{fontenc}
\usepackage{slashed}
\usepackage{setspace}
\usepackage{graphicx}
\usepackage{xcolor}
\usepackage[final,          
            colorlinks,     
            linkcolor=purple, 
            citecolor=teal, 
            urlcolor=blue  
            ]{hyperref}

\makeatletter
\DeclareRobustCommand*{\bfseries}{%
  \not@math@alphabet\bfseries\mathbf
  \fontseries\bfdefault\selectfont
  \boldmath
}
\makeatother

\newcommand{\SU}{\ensuremath{\mathrm{SU}}}

\numberwithin{equation}{section}

\begin{document}
\selectlanguage{english}

\title{(Quasi-)Degeneration, Quantum Corrections and Neutrino
  Mixing\footnote{Talk given at the 42\(^\mathrm{th}\) ITEP winter
    school (Feb 11--18 2014), Otradnoe, Moscow, Russia.}}

\preprint{TTP14-013}

\author{\firstname{W.~G.}~\surname{Hollik}}
\email{hollik@kit.edu}
\affiliation{Institut f\"ur Theoretische Teilchenphysik (TTP), Karlsruhe
  Institute for Technology
}


\begin {abstract}
In the case of neutrinos having significant masses to be measured in
direct search, there are sizeable corrections to the PMNS matrix. We show
the universality of loop effects from any new physics sector and discuss
the seesaw-extended MSSM as an example of a non-minimal flavour violating
theory with non-decoupling effects.
\end{abstract}

\maketitle

\section{Introduction}
In the Standard Model of elementary particle physics, neutrinos play a
special role among the fermions: neutrinos are exactly massless and
neutrinos only appear as left-handed fermions. On the one hand,
oscillation experiments suggest non-vanishing neutrino masses
 \cite{Fukuda:1998mi,Ahmad:2002jz}
which at least have to be in the sub-eV regime. On the other hand,
masses of righthanded neutrinos are unconstrained by the symmetries of
the Standard Model and can in principle be much larger than the
electroweak scale \cite{Minkowski:1977sc}.

\paragraph{Neutrino masses and quasi-degeneration}
The absolute mass scale of the light neutrinos, however, is still
unknown. Global fits on neutrino data give very precise measurements of
the mass squared differences \(\Delta m_{ij}^2 = m_i^2-m_j^2\)
\cite{Capozzi:2013csa} (similar results are obtained by other
groups~\cite{Forero:2014bxa, Gonzalez-Garcia:2014bfa}):
\begin{equation}\label{eq:deltas}
\begin{aligned}
\Delta m_{21}^2 &= 7.54^{+0.26}_{-0.22}\times 10^{-5}\,\mathrm{eV}^2,\\
\Delta m_{31}^2 &= 2.45 \pm 0.06\times 10^{-3}\,\mathrm{eV}^2.
\end{aligned}
\end{equation}
Unknown as well is the hierarchy
of the spectrum: the sign of the larger \(\Delta m_{31}^2\) is yet
undetermined and distinguishes between normal (1-2-3) or inverted
(3-1-2) hierarchy.

If the lightest neutrino is sufficiently heavy, direct mass searches as
done by the Karlsruhe Tritium Neutrino Experiment (KATRIN)
\cite{Osipowicz:2001sq} or
will be able to measure it---or set a tight upper bound on the effective
electron neutrino mass
\(m^2_{\nu_\mathrm{e}} = \sum_i \left|U_{\mathrm{e}i}\right|^2
m_i^2\). After three years of measurement a sensitivity
\(m_{\nu_\mathrm{e}} \leq 0.2\,\mathrm{eV}\) is expected \cite{Angrik:2005ep}.

\paragraph{Quark vs.\ lepton mixing}

In the quark sector, flavour mixing is encoded in the
Cabibbo-Kobayashi-Maskawa (CKM) matrix \cite{Cabibbo:1963yz,Kobayashi:1973fv}
whose mixing angles are measured to be small leading to a CKM matrix
being closely the unit matrix. On the other hand, neutrino mixing is
encoded in the leptonic mixing matrix according to Pontecorvo
\cite{Pontecorvo:1957qd}, Maki, Nakagawa and Sakata \cite{Maki:1962mu}
(PMNS) which is strongly non-hierarchical and shows rather arbitrary
mixing. A schematical comparison between both is shown in Fig.~\ref{fig:CKMvsPMNS}.

\begin{figure}
  \[
V_\text{CKM} = \left(\begin{array}{ccc}
        \begin{picture}(10,10)\put(2.5,2.5){\circle*{9.7428}}\end{picture}
        & \begin{picture}(10,10)\put(2.5,2.5){\circle*{2.253}}\end{picture}
        & \begin{picture}(10,10)\put(2.5,2.5){\circle*{.0347}}\end{picture} \\
        \begin{picture}(10,10)\put(2.5,2.5){\circle*{2.252}}\end{picture}
        & \begin{picture}(10,10)\put(2.5,2.5){\circle*{9.7345}}\end{picture}
        & \begin{picture}(10,10)\put(2.5,2.5){\circle*{.410}}\end{picture} \\
        \begin{picture}(10,10)\put(2.5,2.5){\circle*{.0862}}\end{picture}
        & \begin{picture}(10,10)\put(2.5,2.5){\circle*{.0403}}\end{picture}
        & \begin{picture}(10,10)\put(2.5,2.5){\circle*{9.99152}}\end{picture}
\end{array}\right)\;,
\qquad
U_\text{PMNS} = \left(\begin{array}{ccc}
        \begin{picture}(10,10)\put(2.5,2.5){\circle*{8.35}}\end{picture}
        & \begin{picture}(10,10)\put(2.5,2.5){\circle*{5.27}}\end{picture}
        & \begin{picture}(10,10)\put(2.5,2.5){\circle*{1.57}}\end{picture} \\
        \begin{picture}(10,10)\put(2.5,2.5){\circle*{5.04}}\end{picture}
        & \begin{picture}(10,10)\put(2.5,2.5){\circle*{6.17}}\end{picture}
        & \begin{picture}(10,10)\put(2.5,2.5){\circle*{6.04}}\end{picture} \\
        \begin{picture}(10,10)\put(2.5,2.5){\circle*{2.21}}\end{picture}
        & \begin{picture}(10,10)\put(2.5,2.5){\circle*{5.84}}\end{picture}
        & \begin{picture}(10,10)\put(2.5,2.5){\circle*{7.81}}\end{picture}
\end{array}\right)
\]
  \caption{Sizes of the mixing matrix elements---CKM versus PMNS
mixing.}
  \label{fig:CKMvsPMNS}
\end{figure}

The small mixings of the quark sector suggest to generate them as an
effect of quantum corrections, where the mixing pattern of the PMNS
matrix is usually addressed by postulating flavour symmetries. However,
the possibility of radiative generation of neutrino mixing becomes
important in a regime where neutrino masses are quasi-degenerate which
will be proven by the direct mass searches.

\section{Radiative Flavour Violation}

In the Standard Model, flavour violation is a property of the Yukawa
couplings of fermions to the Higgs sector. Any theory beyond the
Standard Model either provides \emph{minimal flavour violation}, where
all sources of flavour violation are in the Yukawa couplings, or
additional sources of flavour mixing that are tightly constrained by
flavour observables.

\begin{figure}[t]
\begin{minipage}{.5\textwidth}
  \includegraphics[width=\textwidth]{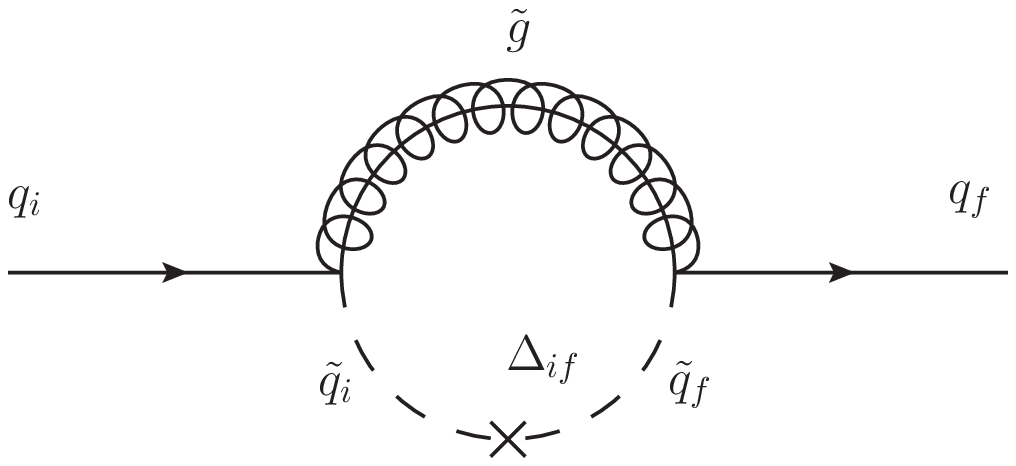}
\end{minipage}%
\begin{minipage}{.5\textwidth}
  \includegraphics[width=\textwidth]{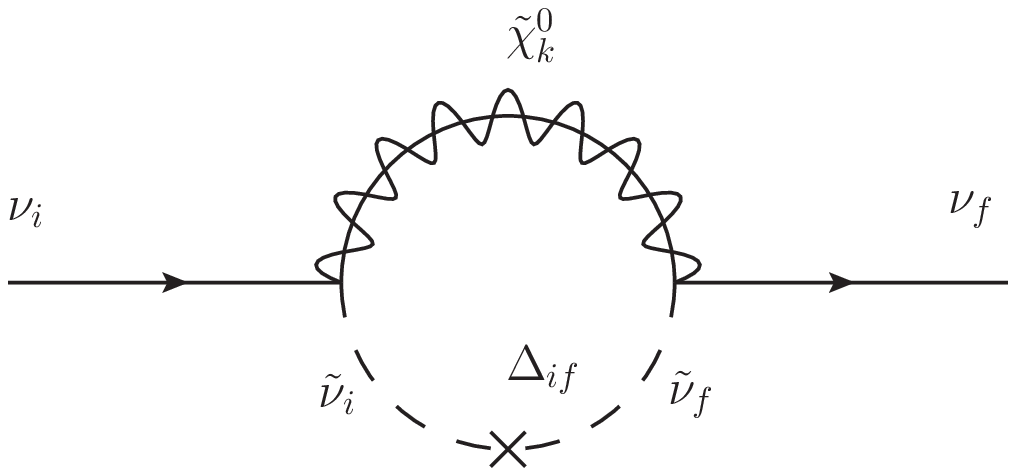}
\end{minipage}
  \caption{Flavour changing self energy by the virtue of SUSY
corrections for quarks (left) and neutrinos (right).}
  \label{fig:fcselfen}
\end{figure}

The general minimal supersymmetric Standard Model (MSSM) carries in
principle arbitrary flavour structures in the terms that softly break
supersymmetry. Soft breaking scalar masses as well as trilinear
couplings of scalar superpartners to the Higgs scalars can mix flavour
in a way that is not accounted for in the Yukawa couplings. Those
contributions enter via the sfermionic mass matrix and induce flavour
changes in self-energies by supersymmetric (SUSY) quantum corrections as
shown in Fig.~\ref{fig:fcselfen}.

\begin{figure}[b]
  \includegraphics[width=\textwidth]{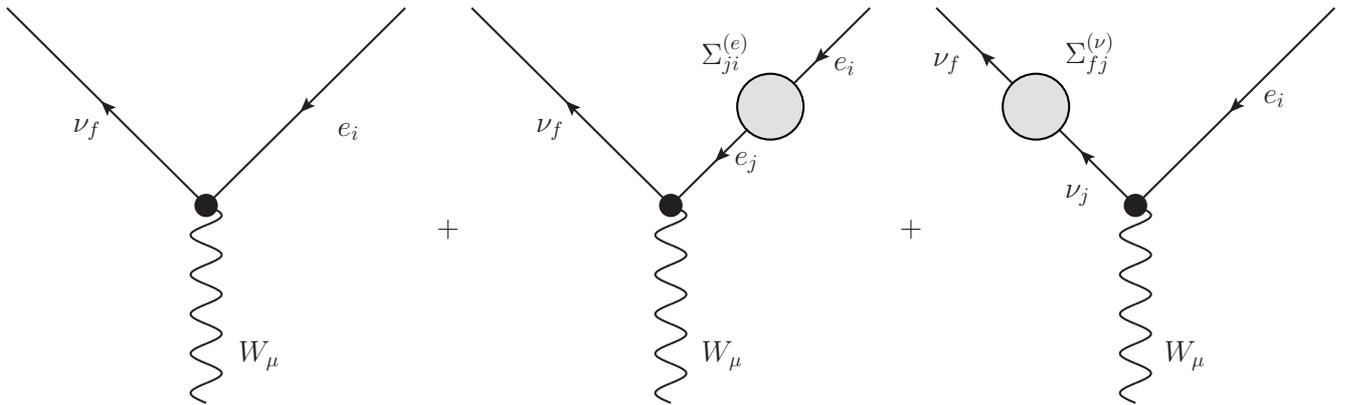}
  \caption{Renormalization of the leptonic mixing matrix by flavour
changing self-energies at external legs.}
  \label{fig:PMNSren}
\end{figure}

Appending those self-energies to charged current interactions (see
Fig.~\ref{fig:PMNSren} for the leptonic case), the flavour mixing matrix
associated with that interaction befalls renormalization according to
\cite{Denner:1990yz}. It was shown that in the MSSM such corrections are
sufficient to generate quark mixing radiatively \cite{Ferrandis:2004ri,
  Crivellin:2008mq}. Moreover, finite supersymmetric threshold
corrections have the power to generate fermion masses
radiatively~\cite{Lahanas:1982et, Masiero:1983ph, Banks:1987iu,
  Borzumati:1999sp}. Large effects from supersymmetry are also known in
lepton flavour violation~\cite{Borzumati:1986qx}.

The renormalized electron-neutrino-W vertex can then be written as
\begin{equation}\label{eq:PMNSren}
i \frac{g}{\sqrt{2}} \gamma^\mu P_L U_\text{PMNS}^\dag
\rightarrow
i \frac{g}{\sqrt{2}} \gamma^\mu P_L \left(
  U^{(0)\dag} + U^{(0)\dag} \Delta U^e + \Delta U^\nu U^{(0)\dag}\right),
\end{equation}
with \(U^{(0)}\) being the unrenormalized mixing matrix. The correction
from the neutrino leg \(\Delta U^\nu\) is sensitive to the mass
spectrum and the flavour-changing self-energy \(\Sigma^\nu_{ij}\)
\begin{equation}
\Delta U^\nu_{fi} \sim \frac{m_{\nu_f} \Sigma^\nu_{fi}}{\Delta m_{fi}^2}
\end{equation}
and the charged lepton contribution \(\Delta U^e\) small in contrast
\cite{Girrbach:2009uy}.

\begin{figure}
  \includegraphics[width=\textwidth]{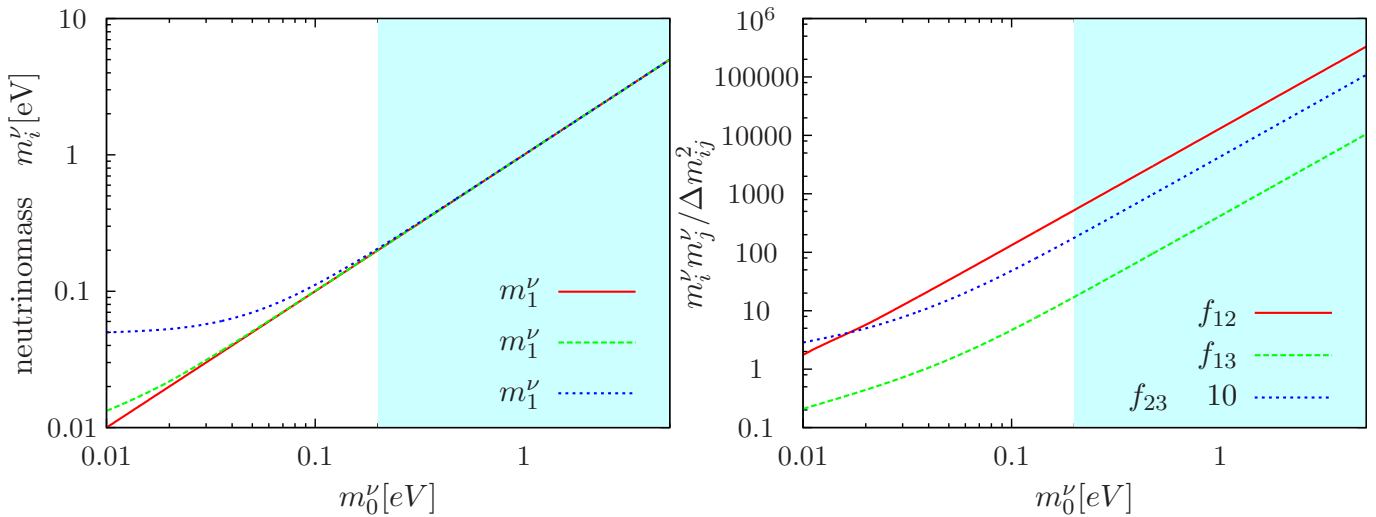}
  \caption{Enhancement factors of radiative corrections to neutrino
mixing (right). The shaded area shows the KATRIN discovery regime.}
  \label{fig:enhancement}
\end{figure}

In the quasi-degenerate regime of neutrino masses, there occurs a strong
enhancement of this type of corrections that can significantly supersede
loop suppression factors if the absolute neutrino mass is around the
KATRIN discovery limit \(m_\nu^{(0)} \approx 0.35\,\mathrm{eV}\), Fig.~\ref{fig:enhancement}:
\[
f_{ij} = \frac{m_{\nu_f} m_{\nu_i}}{\Delta m_{fi}^2} \lesssim 5\times 10^5.
\]

\section{The MSSM with Righthanded Neutrinos}

The standard mass generation mechanism for fermions would force neutrino
Yukawa couplings being orders of magnitude smaller than the others. In
contrast, the seesaw mechanism provides an effective and elegant way to
cope with light neutrinos by keeping their Dirac masses \(m_D\) at the
electroweak scale. A rough estimate of scales sets the mass of
righthanded neutrinos \(m_R\) around \(10^{13\ldots 14}\,\mathrm{GeV}\)
for \(m_D = \mathcal{O}(m_t)\):
\[
m_{\nu_\ell} \sim m_D^2/m_R \approx \mathcal{O}(0.1\,\mathrm{eV}).
\]

The implementation of this simplest type of seesaw mechanism in the MSSM
involves the addition of righthanded neutrino superfields: chiral
superfields that are gauge singlets under the Standard Model group,
coupled to left-handed \(\SU(2)_L\) doublets via a Yukawa-type
interaction and that have Majorana-type masses in the superpotential:
\begin{equation}\label{eq:nuMSSMsuperpot}
\mathcal{W}^\ell = \mu H_d \cdot H_u
- Y_\ell^{IJ} H_d \cdot L_L^I E_R^J + Y_\nu^{IJ} H_u \cdot L_L^I N_R^J +
\frac{1}{2} m_R^{IJ} N_R^I N_R^J,
\end{equation}
with \(L_L = (\ell_L, \tilde\ell_L) \in \SU(2)_L\) and \(E_R = (e_L^c,
\tilde e_R^*)\), \(N_R = (\nu_L^c, \tilde\nu_R^*)\). The dot-product
denotes the \(\SU(2)\)-invariant multiplication. Summation over
\(I\) and \(J\) is understood, for symmetry reasons we take the same
number of \(N_R\) fields as \(L_L\).

To softly break supersymmetry soft breaking terms have to be included in
the sneutrino sector as well:
\begin{equation}\label{eq:Vsoft}
\begin{aligned}
\mathcal{V}_\text{soft} =\;&
(\mathcal{M}_{\tilde\ell}^2)^{IJ} \tilde L_L^{I*} \tilde L_L^J +
(\mathcal{M}_{\tilde e}^2)^{IJ} \tilde e_R^{I} \tilde e_R^{J*} +
(\mathcal{M}_{\tilde\nu}^2)^{IJ} \tilde \nu_R^{I} \tilde \nu_R^{J*} \\
\; -
\big[ & (B_\nu)^{IJ} \tilde \nu_R^{I*} \tilde \nu_R^{J*} +
  A_e^{IJ} H_1 \cdot \tilde L_L^I \tilde e_R^{J*} -
  A_\nu^{IJ} H_2 \cdot \tilde L_L^I \tilde \nu_R^{J*} +
  \text{h.c.}
\big].
\end{aligned}
\end{equation}

\section{Some Results}

The SUSY one-loop corrections according to the flavour changing
self-energies in Fig.~\ref{fig:PMNSren} are dependent on the soft
breaking parameters introduced in Eq.~\ref{eq:Vsoft}, besides the
traditional MSSM parameters: \(\Delta U^\nu = \Delta
U^\nu(\mathcal{M}^2_{\tilde\ell}, \mathcal{M}^2_{\tilde\nu}, A_\nu)\).

Radiative lepton decays \(\ell_j\to\ell_i\gamma\) severely constrain the
off-diagonal elements of the soft breaking mass
\(\mathcal{M}^2_{\tilde\ell}\) as well as the left-right mixing of
charged sleptons, basically \(A_e\). To comply with those flavour
observables, we set them to zero in the numerical example below:
\(A_e\equiv 0\) and \(\mathcal{M}^2_{\tilde\ell} = \mathcal{M}^2_{\tilde
  e} = \mathcal{M}^2_{\tilde\nu} \equiv m_\mathrm{soft}^2
\mathds{1}\). The absolute mass scale for the light neutrinos lies at
the KATRIN discovery limit: \(m_\nu^{(0)} = 0.35\,\mathrm{eV}\).

\begin{figure}
  \includegraphics[width=\textwidth]{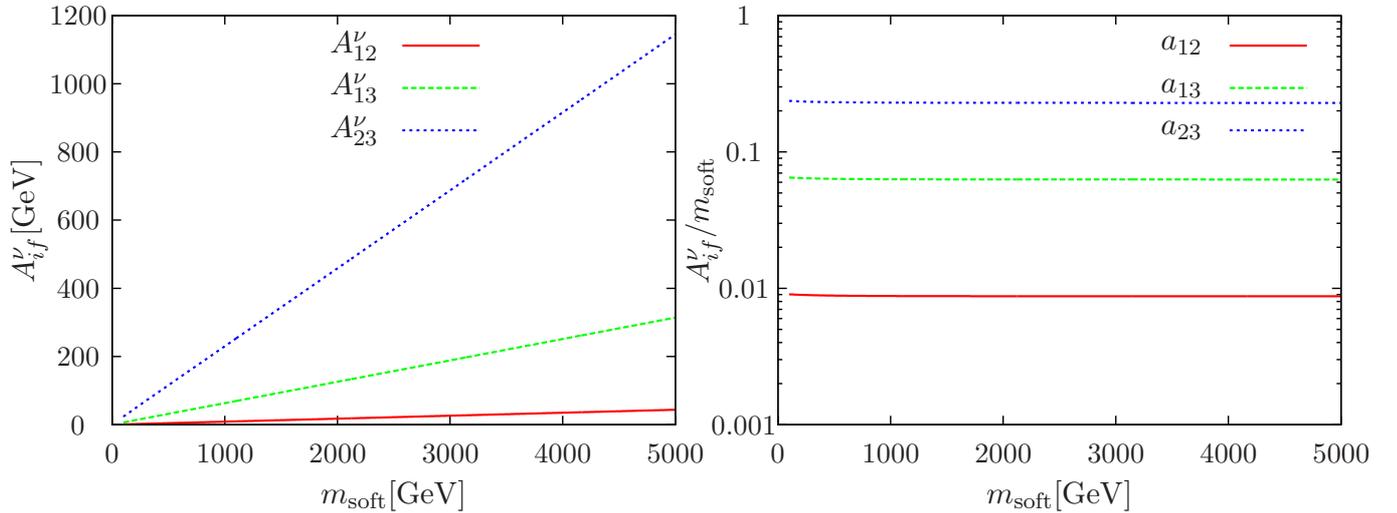}
  \caption{Off-diagonal neutrino \(A\) terms to radiatively generate
PMNS mixing. As can be seen from the right-hand side, this effect does
not decouple with increasing SUSY masses.}
  \label{fig:result}
\end{figure}

To show the power of the method, we artificially suppress potential
flavour mixing in the neutrino Yukawa coupling by alignment
to the charged lepton Yukawa coupling which always can be rotated into a
flavour diagonal basis without affecting the charged current mixing (in
the superpotential of Eq.~\eqref{eq:nuMSSMsuperpot}). The only flavour
violation which is left sits in the neutrino \(A\)
terms---Fig.~\ref{fig:result} shows the size of the off-diagonal
\(A^\nu_{ij}\) needed to produce the corresponding PMNS mixing
\begin{equation}
|U_{12}| \approx 0.52, \qquad
|U_{13}| \approx 0.15, \qquad
|U_{23}| \approx 0.58
\end{equation}
correctly.

\section{Conclusions}

As has been shown in this talk, there are radiative corrections to the
leptonic mixing matrix which get sizeable if neutrinos are nearly
degenerate in masses. We have not included in the analysis the influence
of those corrections to the masses themselves. However, since the
generic structure is not limited to supersymmetric corrections but only
depends on the size of the flavour changing neutrino self-energies
\(\Sigma^\nu_{ij}\) the description can be generalized to any theory
with new flavour structures allowing for non-vanishing
\(\Sigma^\nu_{ij}\) whith \(i\neq j\). In the case of MSSM extensions,
the corrections do not decouple with increasing size of the SUSY
scale. To reduce the influence on flavour changing branching ratios
\(\ell_j\to\ell_i\gamma\) a specific restriction on the SUSY parameter
was taken. However, for the numerical example given in this talk
\(\mathrm{BR}(\mu\to e\gamma) \lesssim 10^{-28}\) complying with data
and similarly other Lepton Flavour Violation constraints are obeyed as
well. The evaluation of the branching ratios was performed according to
the results of \cite{Hisano:1995cp}.

\subsection*{Acknowledgements}
I want to thank Ulrich Nierste for collaborating in this project and
proofreading the manuscript. This work as well as the journey to
Moscow was supported by the Graduiertenkolleg 1694
``Elementarteilchenphysik bei h\"ochster Energie und h\"ochster
Pr\"azision''. Moreover, it is a pleasure to thank the organizers of the
school for their hospitality, setting the environment and skiing lectures.

\bibliographystyle{utcaps}
\bibliography{ITEP_arxiv_wghollik}

\providecommand{\href}[2]{#2}\begingroup\raggedright\begin{thebibliography}{10}

\bibitem{Fukuda:1998mi}
{\bfseries Super-Kamiokande Collaboration} Collaboration, Y.~Fukuda {\em
  et~al.}, ``{Evidence for oscillation of atmospheric neutrinos},''
  \href{http://dx.doi.org/10.1103/PhysRevLett.81.1562}{{\em Phys.Rev.Lett.}
  {\bfseries 81} (1998) 1562--1567},
\href{http://arxiv.org/abs/hep-ex/9807003}{{\ttfamily arXiv:hep-ex/9807003
  [hep-ex]}}.

\bibitem{Ahmad:2002jz}
{\bfseries SNO Collaboration} Collaboration, Q.~Ahmad {\em et~al.}, ``{Direct
  evidence for neutrino flavor transformation from neutral current interactions
  in the Sudbury Neutrino Observatory},''
  \href{http://dx.doi.org/10.1103/PhysRevLett.89.011301}{{\em Phys.Rev.Lett.}
  {\bfseries 89} (2002) 011301},
\href{http://arxiv.org/abs/nucl-ex/0204008}{{\ttfamily arXiv:nucl-ex/0204008
  [nucl-ex]}}.

\bibitem{Minkowski:1977sc}
P.~Minkowski, ``{\(\mu \to e \gamma\) at a Rate of One Out of 1-Billion Muon
  Decays?},''
\href{http://dx.doi.org/10.1016/0370-2693(77)90435-X}{{\em Phys.Lett.}
  {\bfseries B67} (1977) 421}.

\bibitem{Capozzi:2013csa}
F.~Capozzi, G.~Fogli, E.~Lisi, A.~Marrone, D.~Montanino, {\em et~al.},
  ``{Status of three-neutrino oscillation parameters, circa 2013},''
  \href{http://dx.doi.org/10.1103/PhysRevD.89.093018}{{\em Phys.Rev.}
  {\bfseries D89} (2014) 093018},
\href{http://arxiv.org/abs/1312.2878}{{\ttfamily arXiv:1312.2878 [hep-ph]}}.

\bibitem{Forero:2014bxa}
D.~Forero, M.~Tortola, and J.~Valle, ``{Neutrino oscillations refitted},''
\href{http://arxiv.org/abs/1405.7540}{{\ttfamily arXiv:1405.7540 [hep-ph]}}.

\bibitem{Gonzalez-Garcia:2014bfa}
M.~Gonzalez-Garcia, M.~Maltoni, and T.~Schwetz, ``{Updated fit to three
  neutrino mixing: status of leptonic CP violation},''
\href{http://arxiv.org/abs/1409.5439}{{\ttfamily arXiv:1409.5439 [hep-ph]}}.

\bibitem{Osipowicz:2001sq}
{\bfseries KATRIN Collaboration} Collaboration, A.~Osipowicz {\em et~al.},
  ``{KATRIN: A Next generation tritium beta decay experiment with sub-eV
  sensitivity for the electron neutrino mass. Letter of intent},''
\href{http://arxiv.org/abs/hep-ex/0109033}{{\ttfamily arXiv:hep-ex/0109033
  [hep-ex]}}.

\bibitem{Angrik:2005ep}
{\bfseries KATRIN Collaboration} Collaboration, J.~Angrik {\em et~al.},
  ``{KATRIN design report 2004},''
{\em FZKA-7090} (2005) .

\bibitem{Cabibbo:1963yz}
N.~Cabibbo, ``{Unitary Symmetry and Leptonic Decays},''
\href{http://dx.doi.org/10.1103/PhysRevLett.10.531}{{\em Phys.Rev.Lett.}
  {\bfseries 10} (1963) 531--533}.

\bibitem{Kobayashi:1973fv}
M.~Kobayashi and T.~Maskawa, ``{CP Violation in the Renormalizable Theory of
  Weak Interaction},''
\href{http://dx.doi.org/10.1143/PTP.49.652}{{\em Prog.Theor.Phys.} {\bfseries
  49} (1973) 652--657}.

\bibitem{Pontecorvo:1957qd}
B.~Pontecorvo, ``{Inverse beta processes and nonconservation of lepton
  charge},''
{\em Sov.Phys.JETP} {\bfseries 7} (1958) 172--173.

\bibitem{Maki:1962mu}
Z.~Maki, M.~Nakagawa, and S.~Sakata, ``{Remarks on the unified model of
  elementary particles},''
\href{http://dx.doi.org/10.1143/PTP.28.870}{{\em Prog.Theor.Phys.} {\bfseries
  28} (1962) 870--880}.

\bibitem{Denner:1990yz}
A.~Denner and T.~Sack, ``{Renormalization of the Quark Mixing Matrix},''
\href{http://dx.doi.org/10.1016/0550-3213(90)90557-T}{{\em Nucl.Phys.}
  {\bfseries B347} (1990) 203--216}.

\bibitem{Ferrandis:2004ri}
J.~Ferrandis and N.~Haba, ``{Supersymmetry breaking as the origin of flavor},''
  \href{http://dx.doi.org/10.1103/PhysRevD.70.055003}{{\em Phys.Rev.}
  {\bfseries D70} (2004) 055003},
\href{http://arxiv.org/abs/hep-ph/0404077}{{\ttfamily arXiv:hep-ph/0404077
  [hep-ph]}}.

\bibitem{Crivellin:2008mq}
A.~Crivellin and U.~Nierste, ``{Supersymmetric renormalisation of the CKM
  matrix and new constraints on the squark mass matrices},''
  \href{http://dx.doi.org/10.1103/PhysRevD.79.035018}{{\em Phys.Rev.}
  {\bfseries D79} (2009) 035018},
\href{http://arxiv.org/abs/0810.1613}{{\ttfamily arXiv:0810.1613 [hep-ph]}}.

\bibitem{Lahanas:1982et}
A.~Lahanas and D.~Wyler, ``{Radiative Fermion Masses and Supersymmetry},''
\href{http://dx.doi.org/10.1016/0370-2693(83)90696-2}{{\em Phys.Lett.}
  {\bfseries B122} (1983) 258}.

\bibitem{Masiero:1983ph}
A.~Masiero, D.~V. Nanopoulos, and K.~Tamvakis, ``{Radiative Fermion Masses in
  Supersymmetric Theories},''
\href{http://dx.doi.org/10.1016/0370-2693(83)90176-4}{{\em Phys.Lett.}
  {\bfseries B126} (1983) 337}.

\bibitem{Banks:1987iu}
T.~Banks, ``{Supersymmetry and the Quark Mass Matrix},''
\href{http://dx.doi.org/10.1016/0550-3213(88)90222-2}{{\em Nucl.Phys.}
  {\bfseries B303} (1988) 172}.

\bibitem{Borzumati:1999sp}
F.~Borzumati, G.~R. Farrar, N.~Polonsky, and S.~D. Thomas, ``{Soft Yukawa
  couplings in supersymmetric theories},''
  \href{http://dx.doi.org/10.1016/S0550-3213(99)00328-4}{{\em Nucl.Phys.}
  {\bfseries B555} (1999) 53--115},
\href{http://arxiv.org/abs/hep-ph/9902443}{{\ttfamily arXiv:hep-ph/9902443
  [hep-ph]}}.

\bibitem{Borzumati:1986qx}
F.~Borzumati and A.~Masiero, ``{Large Muon and electron Number Violations in
  Supergravity Theories},''
\href{http://dx.doi.org/10.1103/PhysRevLett.57.961}{{\em Phys.Rev.Lett.}
  {\bfseries 57} (1986) 961}.

\bibitem{Girrbach:2009uy}
J.~Girrbach, S.~Mertens, U.~Nierste, and S.~Wiesenfeldt, ``{Lepton flavour
  violation in the MSSM},''
  \href{http://dx.doi.org/10.1007/JHEP05(2010)026}{{\em JHEP} {\bfseries 1005}
  (2010) 026},
\href{http://arxiv.org/abs/0910.2663}{{\ttfamily arXiv:0910.2663 [hep-ph]}}.

\bibitem{Hisano:1995cp}
J.~Hisano, T.~Moroi, K.~Tobe, and M.~Yamaguchi, ``{Lepton flavor violation via
  right-handed neutrino Yukawa couplings in supersymmetric standard model},''
  \href{http://dx.doi.org/10.1103/PhysRevD.53.2442}{{\em Phys.Rev.} {\bfseries
  D53} (1996) 2442--2459},
  \href{http://arxiv.org/abs/hep-ph/9510309}{{\ttfamily arXiv:hep-ph/9510309
  [hep-ph]}}.

\end{thebibliography}\endgroup

\end{document}